\documentclass[pra, aps, twocolumn]{revtex4}

\usepackage{slashed}
\usepackage{graphicx}
\usepackage{subfigure}
\usepackage[usenames, dvipsnames]{color}
\usepackage{graphics}
\usepackage{hyperref}
\usepackage{bm}
\usepackage{amsfonts}
\hypersetup{backref,
colorlinks=true,
linkcolor=blue,
linktoc=page,
citecolor=blue}

\begin{document}

\title{Exciton-polariton wakefields in semiconductor microcavities}

\author{H. Ter\c cas}

\email{htercas@gmail.com}

\affiliation{Institut Pascal, PHOTON-N2, Clermont Unversit\'e, Blaise Pascal University, CNRS, 24 Av. des Landais, 63177 Aubiere Cedex, France}

\author{J. T. Mendon\c{c}a}

\email{titomend@ist.utl.pt}

\affiliation{Instituto de F\'isica, Universidade de S\~ao Paulo, S\~ao Paulo SP, 05508-090 Brasil
\\ IPFN, Instituto Superior T\'ecnico, 1049-001 Lisboa, Portugal}

\begin{abstract}

We consider the excitation of polariton wakefields due to a propagating source in a semiconductor micro cavity. We show that two kinds of wakes are possible, depending on the constituents fraction (either exciton or photon) of the polariton wavefunction. The nature of the wakefields (pure excitonic or polaritonic) can be controled by changing the speed of propagation of the external pump. This process could be used as a diagnostic for the internal parameters of the microcavity.

\end{abstract}

\maketitle

\section{Introduction}

Semi-conductor microcavities, designed to increase the light-matter coupling, consist of a pair of distributed Bragg mirrors confining and electromagnetic mode and one (or several) quantum wells with an exciton resonance, which are placed at the antinodes of the electric field \cite{review, carusotto}. In the strong coupling regime, where coherent exciton-photon interaction overcomes the damping caused by the finite life-time of excitons and cavity photons, a new type of elementary excitations, called exciton polaritons (or cavity polariton), appears in the system \cite{hopfield}. Polaritons are therefore a mixture of the semi-conductor excitations (excitons) with light (photons). \par

One of the crucial aspects for the rapid development of research in the field of semi-conductor microcavities stems in the fact of polaritons may undergo Bose-Einstein condensation \cite{malpuech, zamfirescu} - which has been experimentally  observed in a series of works \cite{kasprzak, richard, balili, utsunomiya, baumberg, kasprzak2, wertz} -, putting together the fields of quantum optics and Bose-Einstein condensates (BEC) \cite{pitaevskii}. The most important differences with respect to the usual atomic BECs are i) the possibility of condensation to occur at higher temperatures, as a consequence to the very small polariton mass (typically, $m\sim 10^{-5}m_e$) and ii) the fact of polariton BEC being a driven-dissipative process \cite{wouters, chiocchetta}, where the condensed steady-state results from the balance of photon and exciton losses with an external pump source. Nevertheless, it has also been shown that a thermodynamic regime for polariton condensates created by a non-resonant pump (i.e. pumped far detuned with respect to the condensation energy minimum) is possible, allowing for the definition of a temperature and a chemical potential for the system \cite{kasprzak2, wertz}. Another important property of polariton BECs concerns superfluidity. The phase transition expected for two-dimensional polaritons is rather a Berezinskii-Kosterlitz-Thouless (BKT) transition toward a superfluid state \cite{malpuech2} and not a true BEC. Such a phase transition has not been immediately observed in CdTe-based and GaN- based structures, because of the presence of a strong structural disorder, which leads to the formation of a glass phase\cite{malpuech3}, or to condensation in a potential minimum of the disorder potential \cite{sanvitto}. Signatures of BKT transition have nevertheless been reported in cleaner, GaAs-based, samples \cite{lai}. 
Some interesting features related to the nonlinearity of the system, such as amplification \cite{savvidis} and optical bistability \cite{baas} have been investigated. Moreover, recent theoretical studies and experimental observation of topologically stable half-solitons \cite{flayac, solnyshkov, hivet} and half-vortices \cite{rubo, manni} in spinor polariton condensates, has allowed to study the dynamics and many-body properties of topological defects in the presence of external fields \cite{tercas, tercas2, tercas3}, thus motivating the exploration of condensed-matter physics with polaritons.   

In the most usual experimental configurations, the external pump is fixed, occupying a well defined region of the planar cavity. However, if we allow the pump to move, the occurrence of new time-dependent phenomena can be expected in exciton-polaritons, even below the condensation threshold. In this work, we study the properties wakefields in semi-conductor microcavities excited by a pump moving with constant velocity. Wakefields are universal phenomena which can be produced by the motion of a boat in the surface of a lake, or by a laser pulse propagating in a gas, having important technological implications in the case of laser-plasma acceleration \cite{tajima}. When an intense electromagnetic pulse hits the plasma, it produces a wake of plasma oscillations through the action of a ponderomotive force. Electrons trapped in the wake can then be accelerated up to very high (relativistic) energies, providing an alternative yet efficient way of accelerating charged particles  \cite{bob, book1}. Acoustic wakefields produced by a Bose-Einstein condensate moving across a thermal (non-condensed) gas has also been considered \cite{becwake, book2}.  Recently, wakefield excitation in metallic nanowires has also been investigated and pointed out as mechanism to produce energetic ultra-violet (XUV) radiaton \cite{ali}. In the present work, we show that a similar process can occur in a gas of  excitons and exciton-polaritons (we should refer to the latter as ``polaritons"), depending on the point of the dispersion that the system is pumped. 

The structure of this paper is the following. In Section II, we establish the basic equations of our problem. We start from the coupled photon-exciton wave equations and derive the dispersion relations of the two polariton branches, which are the basic excitations of the electromagnetic field coupled with the excitonic mode. We then include external pump term and derive the appropriate wakefield equations. In Section III, we derive the wakefield solution for an exciton gas, by assuming that the lower polariton branch is pumped in the exciton-dominated part (high-wavevector). In Section IV, we derive the general form of polariton wakefields, produced if the pumped is tuned near the bottom of the polariton dispersion. Finally, in Section V, we state some conclusions.

\section{Basic equations}

The coupled dynamics of the photonic and excitonic fields, represented by $\phi ({\bf r}, t)$ and $\chi ({\bf r}, t)$ respectively, can be described by the following system of wave equations \cite{review, shelykh2, flayac3}
\begin{equation}
i \hbar \frac{\partial \phi}{\partial t} = - \left[ \frac{\hbar^2}{2 m_\phi} \nabla^2 + \frac{i \hbar}{2 \tau_\phi} \right] \phi + \frac{\hbar}{2} \Omega_R \chi + P,
\label{2.1a} 
\end{equation} 
\begin{equation}
i \hbar \frac{\partial \chi}{\partial t} = - \left[ \frac{\hbar^2}{2 m_\chi} \nabla^2 + \frac{i \hbar}{2 \tau_\chi} \right] \chi + \frac{\hbar}{2} \Omega_R \phi + \alpha_1 \left \vert \chi \right \vert^2 \chi .
\label{2.1b} \end{equation}
Here, $m_\phi \ll m_e$ and $m_\chi \leq m_e$ are the photon and exciton masses, respectively ($m_e$ represents the electron mass), and $\Omega_R$ is the Rabi frequency measuring the strength of the coupling. The nonlinear term $\alpha_1=6E_ba_B^2/\mathcal{S}$, where $\mathcal{S}$ is the normalization surface, $E_b$ is the exciton binding energy and $a_B$ the corresponding Borh radius, accounts for the polariton-polariton contact interactions \cite{ciuti}. Typical values are $m_\phi = 5 \times 10^{-5} m_e$, and $m_\chi = 0.4 m_e$ and $\hbar \Omega_R=10$ meV. The quantities $\tau_\phi$ and $\tau_\chi$ are the lifetimes of cavity photons and excitons, with typical values $\tau_\phi = 10$ ps and $\tau_\chi = 400$ ps. \par
\begin{figure}
\centering
\includegraphics[scale=0.6]{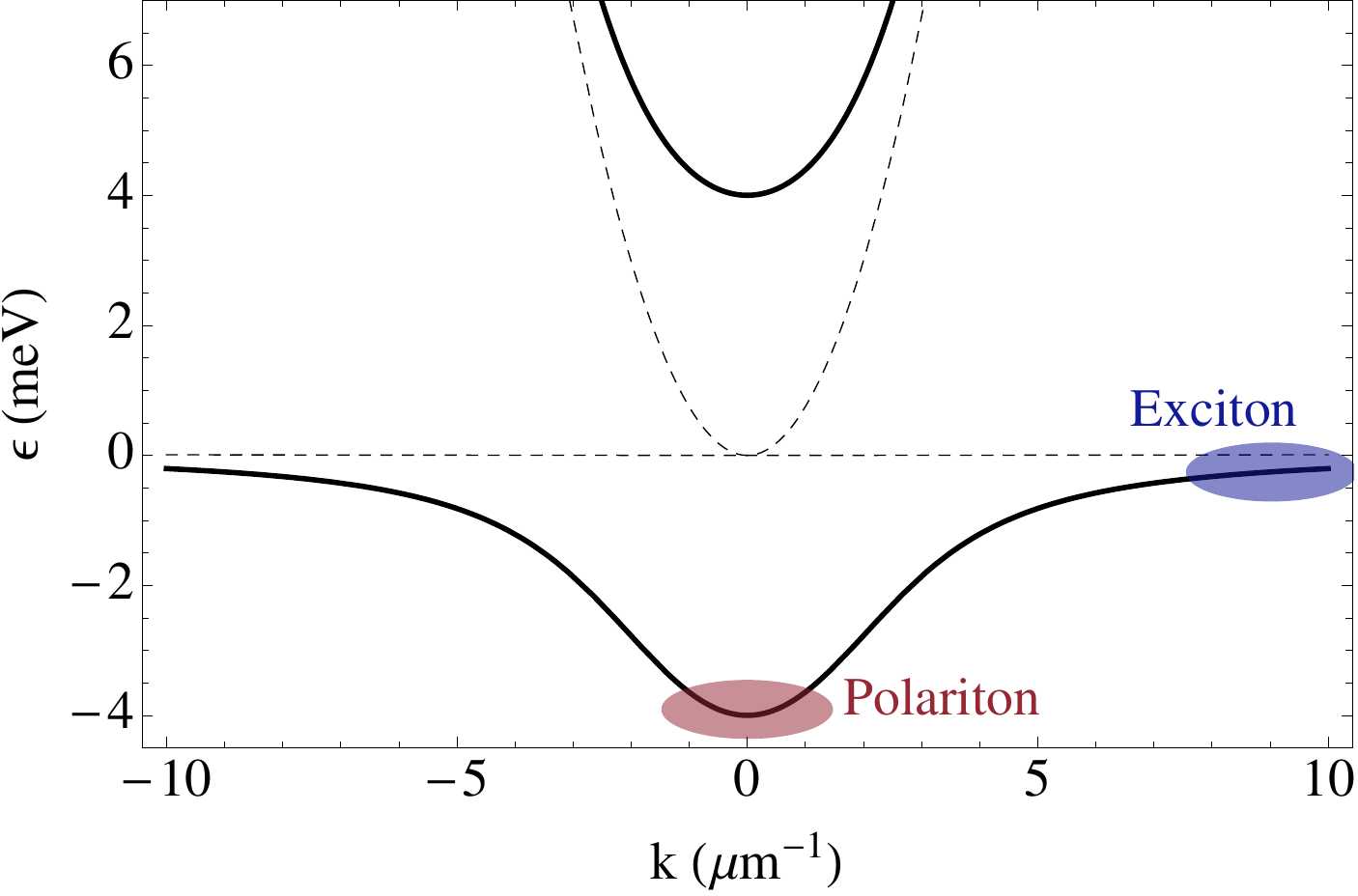}
\caption{Lower and upper polariton branches in a semi-conductor microcavity. We excite different wavevector ranges in the lower branch to generate the wake fields. We have used the parameters for a GaAs microcavity: $m_\phi=5\times 10^{-5}m_e$, $m_\chi=0.4m_e$,  and $\hbar\Omega_R=10$ meV.   }
\label{fig1}
\end{figure}

In Eq. (\ref{2.1a}), $P \equiv P ({\bf r}, t)$ is the pump, acting as a source of photons, and resulting from an external laser beam to be quasi-resonantly tuned with respect to the lower polariton branch, as we specify bellow. We start by deriving the appropriate dispersion relations fro the photon-exciton coupled field. The bare photonic and excitonic modes are readily obtained by neglecting the Rabi coupling $\Omega_R$ and the interaction term $\alpha_1$, 
\begin{equation}
i \hbar \frac{\partial \phi}{\partial t} = - \left[ \frac{\hbar^2}{2 m_\phi} \nabla^2 + \frac{i \hbar}{2 \tau_\phi} \right] \phi,
\label{2.3a}
\end{equation} 
\begin{equation}
i \hbar \frac{\partial \chi}{\partial t} = - \left[ \frac{\hbar^2}{2 m_\chi} \nabla^2 + \frac{i \hbar}{2 \tau_\chi} \right] \chi .
\label{2.3b} \end{equation}
We can then find solutions of the form
\begin{equation}
\begin{array}{c}
\displaystyle{\phi({\bf r}, t) = \phi_0 \exp (i {\bf k} \cdot {\bf r} - i \omega_\phi t ),} \\\\
\displaystyle{\chi ({\bf r}, t) = \chi_0 \exp (i {\bf k} \cdot {\bf r} - i \omega_\chi t ) .}
\label{2.4} 
\end{array}
\end{equation}
Replacing in Eqs. (\ref{2.3a})-(\ref{2.3b}), we obtain the dispersion relations
\begin{equation}
\displaystyle{\omega_\phi = \frac{\hbar k^2}{2 m_\phi} + i \gamma_\phi,} \quad
\displaystyle{\omega_\chi = \frac{\hbar k^2}{2 m_\chi}  + i \gamma_\chi,}
\label{2.5} 
\end{equation}
where two damping rates read $\gamma_\phi = 1 / 2 \tau_\phi$ and $\gamma_\chi = 1 / 2 \tau_\chi$. For a given value of the wavenumber $k$, we have $\omega_\phi \gg \omega_\chi$, as a result of the mass difference $m_\phi \ll m_\chi$. The polariton modes can be obtained for a finite value of the Rabi field. Using solutions of the form
\begin{equation}
\left( \phi, \chi  \right) ({\bf r}, t) = \left( \phi_0 , \chi_0 \right) \exp (i {\bf k} \cdot {\bf r} - i \omega t ) ,
\label{2.6} \end{equation}
we can then derive the dispersion equation
\begin{equation}
(\omega - \omega_\phi) (\omega - \omega_\chi) = \frac{\tilde \Omega_R^2}{4},
\label{2.6b} 
\end{equation}
where $\tilde \Omega_R^2=\Omega_R^2-4\gamma_\phi\gamma_\chi$ and $\omega_\phi$ and $\omega_\chi$ are determined by Eq. (\ref{2.5}). Solving for $\omega$, we get the two solutions $\omega_\pm$, such that
\begin{equation}
\omega_\pm = \frac{1}{2} \left[ (\omega_\phi + \omega_\chi) \pm \sqrt{ (\omega_\phi + \omega_\chi) ^2-4\omega_\phi\omega_\chi + \tilde \Omega_R^2} \right] .
\label{2.7} 
\end{equation}
This corresponds to the well known lower ($\omega_-$) and upper ($\omega_+$) polariton branches, as illustrated in Fig. \ref{fig1}. Near $k=0$, the polariton is composed by 50\% photon and 50\% exciton, while for higher values of $k\gg0$, the excitonic fraction is much higher than the photonic one. Thus, the strategy of producing different kind of wakefields consists in moving the pump with different velocities, corresponding to different points in the lower polariton brunch.   

Let us now introduce a finite pump $P\neq 0$, while keeping the interaction term negligible. This can be done if the pump intensity is lower than the condensation threshold for condensation \cite{malpuechbook}. We further assume that the pump moves with constant velocity $v$ along some direction $z$, and assume it to be uniform in the perpendicular direction, which allows us to write $P ({\bf r}, t) \equiv P(x - v t)$. We notice that an equivalent scheme, where the relative motion between the excitons and the pump $P$ can be generated by creating an exciton flow with a second, homogeneous pump. This allows our analysis to be general and overcome the technical issue of moving the laser spot. By performing the following variable transformation $(x, t) \rightarrow (\xi, \tau)$, with
$\xi = (x - v t) \; , \quad \tau = t$, we can rewrite the non-homogeneous exciton-photon field equations in terms of the Lagrangian variable $\xi$ as
\begin{equation}
\left( \frac{\partial^2}{\partial \xi^2} - i \kappa_\phi \frac{\partial}{\partial \xi} + i \Gamma_\phi \right) \phi = \epsilon_\phi \chi + I (\xi) ,
\label{2.8a} \end{equation}
\begin{equation}
\left( \frac{\partial^2}{\partial \xi^2} - i \kappa_\chi \frac{\partial}{\partial \xi} + i \Gamma_\chi - \alpha \left| \chi \right |^2 \right) \chi = \epsilon_\chi \phi ,
\label{2.8b} \end{equation}
where we have introduced the new quantities
\begin{equation}
\kappa_\phi = \frac{2 m_\phi}{\hbar} v \; , \quad  \Gamma_\phi = \frac{2 m_\phi}{\hbar} \gamma_\chi \; , \quad \epsilon_\phi = \frac{m_\phi}{\hbar} \Omega_R ,
\label{2.9a} \end{equation}
and 
\begin{equation}
\kappa_\chi = \kappa_\phi \frac{m_\chi}{m_\phi}  \; , \quad  \Gamma_\chi = \Gamma_\phi \frac{m_\chi}{m_\phi}  \frac{\tau_\phi}{\tau_\chi} \; , \quad \epsilon_\chi = \epsilon_\phi \frac{m_\chi}{m_\phi} ,
\label{2.9b} \end{equation}
with the new source term $I (\xi)$ and nonlinear parameter $\alpha$ defined as
\begin{equation}
I (\xi) = \frac{2 m_\phi}{\hbar^2} P (\xi) \; , \quad \alpha = \frac{2 m_\chi}{\hbar} \alpha_1 .
\label{2.9c} \end{equation}
Eqs. (\ref{2.8a})-(\ref{2.8b}) are the basic equation for the description of the wakefields produced by the source $I (\xi)$ and are derived by assuming the quasi-static approximation, i.e. $\partial / \partial \tau = 0$. Such an approximation can be justified when the dispersion of the driving pump pulse is negligible, and the shape of the pump field $I (\xi)$ stains unchanged (which is the case for an externally applied pump). 
 
\section{Exciton wakefields}

\begin{figure*}
\includegraphics[scale=0.8]{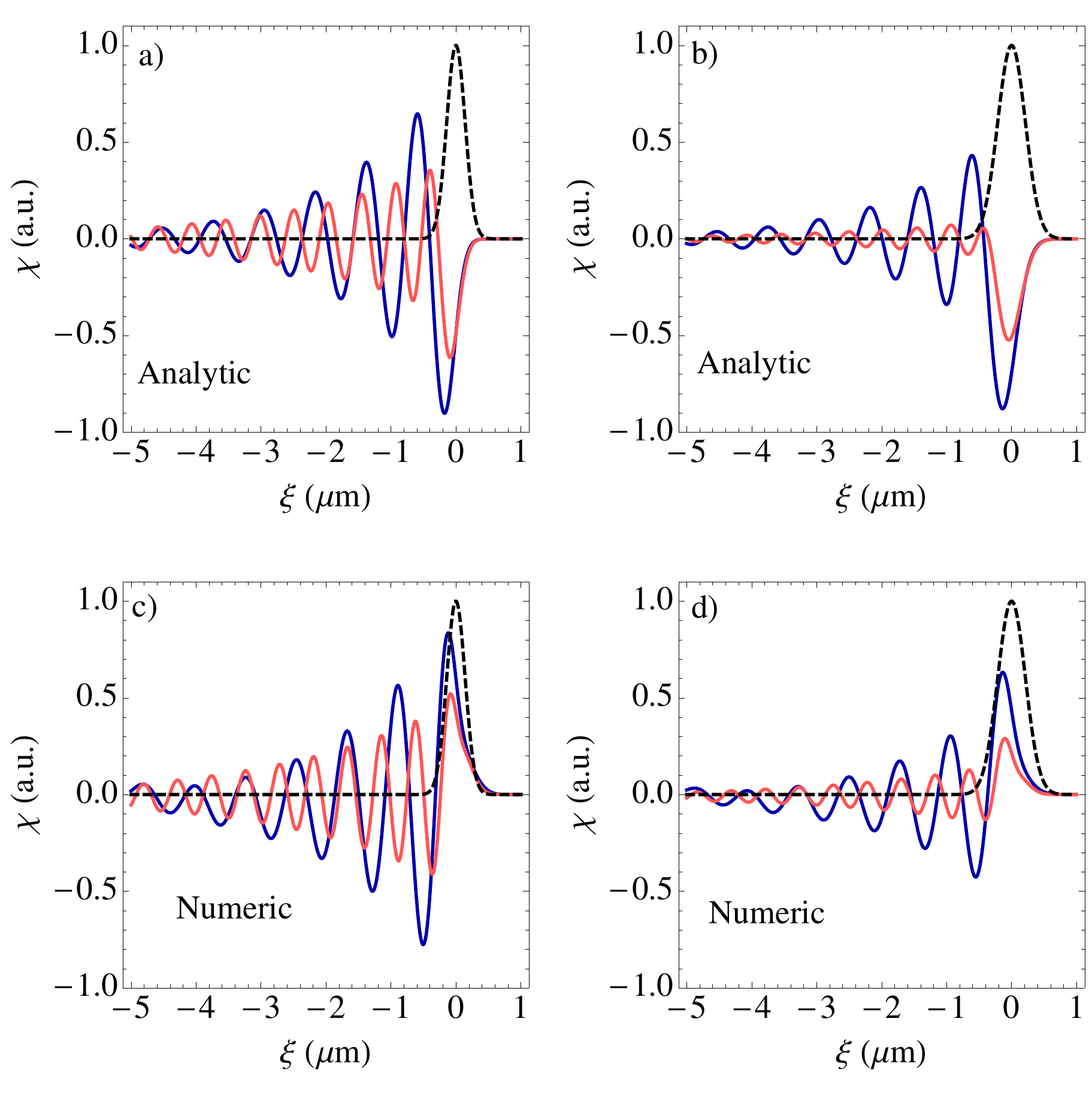}
\caption{(color online) Exciton wakefield produced by an external source (pump) propagating with velocity $v=1.2$ nm/ps (blue/darkgray lines, corresponding to $\kappa_\chi=8.4~\mu$m$^{-1}$) and $v=1.8$ nm/ps (red/light gray lines, corresponding to $\kappa_\chi=12~\mu$m$^{-1}$.  Panels a) and b) are the analytic solutions obtained from Eq. (\ref{3.6}), while panels b) and d) depict the numerical solutions obtained from Eq. (\ref{3.1}). We have used $\sigma_0=0.2~\mu$m for the panels a) and c) (resp. $\sigma_0=0.4~\mu$m for the panels b) and d). In all situations, we have normalized the wakefield amplitude relatively to the pump intensity $I_0$ and used the following parameters \cite{review, malpuechbook}: $m_\phi=5\times 10^{-5}m_e$, $m_\chi=0.4m_e$, $\tau_\phi=30$ ps, $\tau_\chi=400$ ps and $\hbar\Omega_R=10$ meV.  }
\label{fig2}
\end{figure*}

In what follows, we consider the situation where the excitonic fraction of polaritons largely dominates the photon fraction. This corresponds to a large wave-vector propagation, $k\gg0$, corresponding to a pump propagating with velocity $v\sim\hbar k/m_\chi$, as depicted in Fig. \ref{fig1} (blue shadowed region). In this case, we can assume a local photon field solution of the form $i \Gamma_\phi \phi = I (\xi)$, and reduce the above two coupled equations to a single equation for the driven exciton field, of the form
\begin{equation}
\left( \frac{\partial^2}{\partial \xi^2} - i \kappa_\chi \frac{\partial}{\partial \xi} + i \Gamma_\chi - \alpha \left| \chi \right|^2 \right) \chi = - i \frac{\epsilon_\chi}{\Gamma_\chi} I (\xi) .
\label{3.1}
\end{equation}
Let us now define new variables such that
\begin{equation}
\chi (\xi) = h (\xi) \exp \left( \frac{i}{2} \kappa_\chi \xi \right) ,
\label{3.2} \end{equation}
where $h (\xi)$ is the exciton field envelope. Replacing in Eq. (\ref{3.1}), we write 
\begin{equation}
\left[ \frac{\partial^2}{\partial \xi^2} + \kappa^2 (\xi) \right] h = J (\xi) ,
\label{3.3} \end{equation}
where $\kappa (\xi)$ is a slowly varying function of the coordinate $\xi$, as defined by
\begin{equation}
\kappa^2 (\xi) = \frac{\kappa_\chi^2}{4} + i \Gamma_\chi - \alpha | h (\xi) |^2.
\label{3.3b}
\end{equation}
The exciton wakefield equation (\ref{3.3}) describes a driven nonlinear oscillator. For causality reasons, the forced solution is such that the signal $h (\xi)$ must vanish upstream to the source $I (\xi)$, i.e. $h(\xi)=0$ for $\xi>0$. In that case, the formal solution for the exciton wavefunction envelope reads
\begin{equation}
h (\xi) = \int_\infty^\xi J (\xi') \sin \left[ \varphi (\xi) -\varphi (\xi') \right] d \xi' ,
\label{3.4} \end{equation}
with the phase function $\varphi (\xi)$ defined as
\begin{equation}
\varphi (\xi) = \int^\xi \kappa (\xi') d \xi' .
\label{3.4b}
 \end{equation}
Using the expression in Eq. (\ref{3.3b}), and assuming that the term in $\kappa_\chi$ is dominant, which corresponds to a weakly damped and low amplitude wake fields, such that $\kappa_\chi^2 \gg \Gamma_\chi$, and $\kappa_\chi^2 \gg \alpha | h (\xi) |^2$, we get
\begin{equation}
\varphi (\xi) \simeq \frac{1}{2} \left( \kappa_\chi + 2 i \frac{\Gamma_\chi}{\kappa_\chi} \right) \xi - \frac{2 \alpha}{\kappa_\chi} \int^\xi | h (\xi') |^2 d \xi' .
\label{3.5} \end{equation}
In the linear case, when we can assume that $\alpha \simeq 0$, the wakefield solution (\ref{3.4}) reduces to
\begin{equation}
h (\xi) = \int_\infty^\xi J (\xi') \sin \left[ \frac{\kappa_\chi}{2} \left(1 + 2 i \frac{\Gamma_\chi}{\kappa_\chi^2} \right) (\xi - \xi' ) \right] d \xi' .
\label{3.6} 
\end{equation}
These solutions are illustrated in Fig. \ref{fig2}. As we can observe, the wakefield amplitude decreases downstream the pump, as a consequence of the finite life-time of the excitons. Moreover, the amplitude of the wake decreases with the width $\sigma_0$ of the laser.

\section{Polariton wakefields}

Let us now consider that the pump is tuned to be resonant with the bottom of the lower polariton branch, as illustrated in Fig. \ref{fig1}. In that case, where the excitonic and the photonic fractions are nearly equal. We go back to the Eqs. (\ref{2.8a})-(\ref{2.8b}) and defined the new field variables $g (\xi)$ and $h (\xi)$, such that
\begin{equation}
\phi (\xi) = g (\xi) \exp \left( \frac{i}{2} \kappa_\phi \xi \right) \; , \quad
\chi (\xi) = h (\xi) \exp \left( \frac{i}{2} \kappa_\chi \xi \right) .
\label{4.1} \end{equation}
Replacing in the coupled wakefield equations, we get
\begin{equation}\
D_\phi^2 g \equiv \left( \frac{\partial^2}{\partial \xi^2} + \kappa_\phi^2 \right) g = a_\phi h + I (\xi) e^{- i \kappa_\phi \xi / 2} ,
\label{4.2a} \end{equation}
\begin{equation}
D_\chi^2 h \equiv \left[ \frac{\partial^2}{\partial \xi^2} + \kappa_\chi^2 (\xi) \right] h = a_\chi g ,
\label{4.2b} \end{equation}
with the new quantities
\begin{equation}
\kappa_\phi'^2 = \frac{\kappa_\phi^2}{4} + i \Gamma_\phi \; , \quad
\kappa_\chi'^2 (\xi) = \frac{\kappa_\chi^2}{4} + i \Gamma_\chi - \alpha | h (\xi) |^2 ,
\label{4.3} \end{equation}
and
\begin{equation}
a_\phi = \epsilon_\phi e^{i (\kappa_\chi - \kappa_\phi) \xi / 2} \; , \quad
a_\chi = \epsilon_\chi e^{- i (\kappa_\chi - \kappa_\phi) \xi / 2} .
\label{4.4} \end{equation}
We can derive a closed equation for the exciton envelope $h (\xi)$ from Eqs. (\ref{4.2a})-(\ref{4.2b}), which can be written as
\begin{equation}
D_\phi^2 D_\chi^2 h = \frac{\partial^2 \ln a_\chi}{\partial \chi^2} D_\chi^2 h + a_\phi a_\chi h + a_\chi I (\xi) e^{- i \kappa_\phi \xi / 2} .
\label{4.5} \end{equation}
Rearranging the terms, this can finally yield
\begin{equation}
\left( \frac{\partial^4}{\partial \xi^4} + \kappa_a^2 \frac{\partial^2}{\partial \xi^2} + \kappa_b^4 \right) h = J' (\xi) ,
\label{4.6} \end{equation}
where we have used $J' = a_\chi I (\xi) \exp (- i \kappa_\phi \xi / 2)$ and 
\begin{equation}
\kappa_a^2 = \kappa_\phi^2 + \kappa_\chi^2 + (\kappa_\chi - \kappa_\phi )^2 / 4 \; , \quad
\kappa_b^4 = \kappa_\phi^2 \kappa_\chi^2 + (\kappa_\chi - \kappa_\phi )^2 \kappa_\phi^2 / 4 - \epsilon_\phi \epsilon_\chi .
\label{4.7} \end{equation}
Eq. (\ref{4.6}) is the polariton wakefield equation, which includes both the exciton and the photon dynamics. This equation can solved by using the Green's function method. We take the Fourier transform of the latter to obtain
\begin{equation}
h (q) = J' (q) G (q') ,
\label{4.8} 
\end{equation}
where $h (q)$ and $J' (q)$ are the Fourier components of the functions $h (\xi)$ and $J' (\xi)$, as defined by
\begin{equation}
h (\xi) = \int h (q) e^{i q \xi} \frac{d q}{2 \pi} \; , \quad
J' (\xi) = \int J' (q) e^{i q \xi} \frac{d q}{2 \pi} ,
\label{4.8b} \end{equation}
and $G (q)$ is the Fourier transformed Green's function, as determined by
\begin{equation}
G (q) = \frac{1}{(q^4 - q^2 \kappa_a^2 + \kappa_b^4)} .
\label{4.9} \end{equation}
Applying the convolution theorem, we can now write the wakefield solution to Eq. (\ref{4.6}) as
\begin{equation}
h (\xi) = \int_{-\infty}^\infty J' (\xi) G (\xi - \xi') d \xi' ,
\label{4.10} \end{equation}
where $G (\xi)$ is the inverse Fourier transform of $G (q)$. We then get the following solution
\begin{widetext}
\begin{equation}
h (\xi) = \frac{2}{\kappa_a^2} \int_{-\infty}^\infty \left\{ \frac{1}{q_+} \sin [ q_+ (\xi - \xi') ] - \frac{1}{q_-} \sin [ q_- (\xi - \xi') ] \right\} H (\xi - \xi') J' (\xi') d \xi' .
\label{4.11} 
\end{equation}
\end{widetext}
Here, $H (\xi)$ is the Heaviside function, and $q_\pm$ are the two poles of Eq. (\ref{4.9}), 
\begin{equation}
q_\pm^2 = \frac{1}{2} \kappa_a^2 \left( 1\pm \sqrt{1 - 4 \frac{\kappa_b^4}{\kappa_a^4}} \right),
\label{4.12} \end{equation}
which can also be written as
\begin{equation}
q_\pm^2 = \frac{1}{2} \kappa^2 \left( 1\pm \sqrt{1 - 4 \frac{\epsilon_\phi \epsilon_\chi}{\kappa^4}} \right) \; , \quad \kappa^2 = \kappa_\phi^2 + \kappa_\chi^2 + (\kappa_\chi - \kappa_\phi)^2 / 4 .
\label{4.12b} 
\end{equation}
 \begin{figure}[t!]
\includegraphics[scale=0.65]{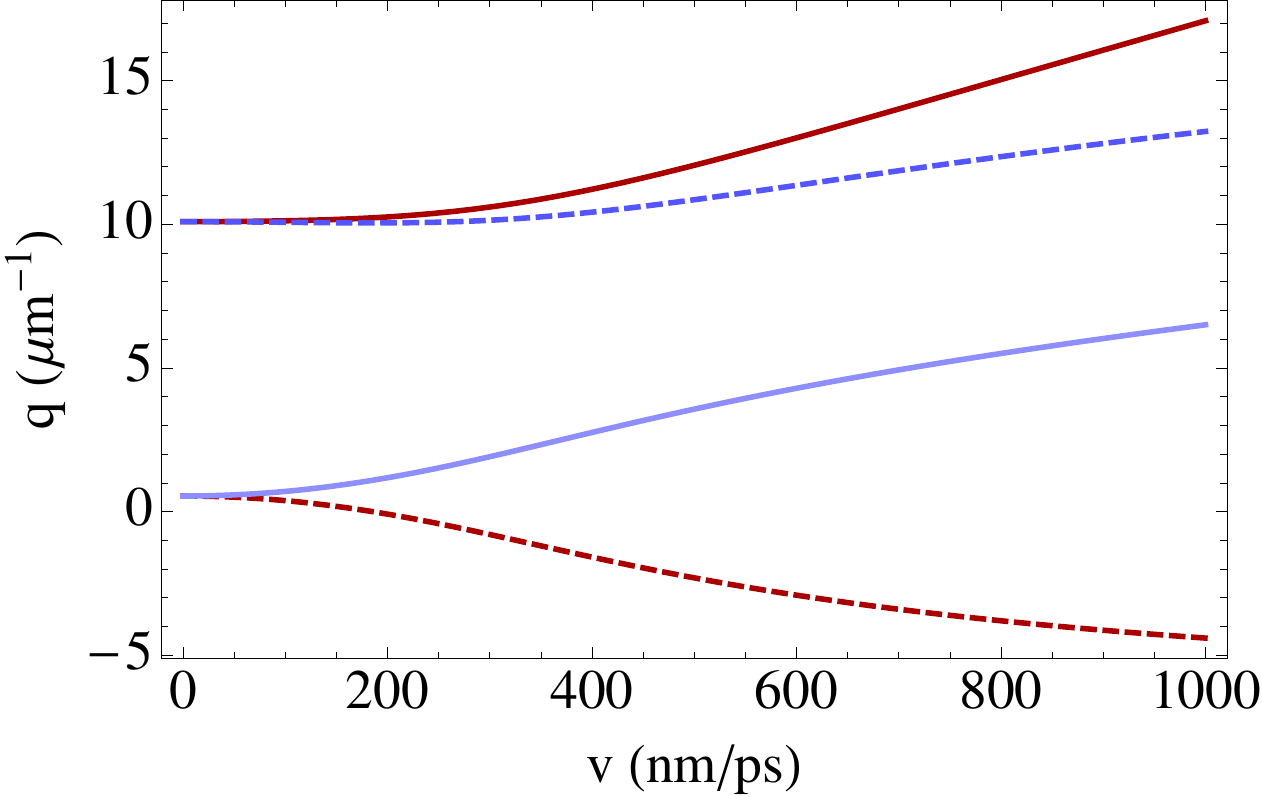}
\caption{(color online) Wakefield wave vectors $q_-$ (blue/light gray) and $q_+$ (red/dark gray) as a function of the pulse velocity $v$. The solid (dashed) lines represents the real (imaginary) parts of $q_-$ and $q_+$. We observe that the fast mode $q_+$ is less damped.}
\label{fig3}
\end{figure}
The slow and fast wake modes $q_\pm$ are separated of, at least, one order of magnitude, as shown in Fig. \ref{fig3}. Similarly to the case of excitonic wakes described in the previous Section, finite exciton (and in the present case also photon) lifetime leads to wake damping, which is related to the imaginary of the modes $q_\pm$. \par

In order to illustrate the physical meaning of the wakefield solution (\ref{4.11}), we determine the wake left behind a Gaussian pump, as defined by $I (\xi) = I_0 \exp (- \xi^2 / \sigma_0^2)$, where $\sigma_0$ is the source width and $I_0 = (2 m_\phi/ \hbar^2) P_0$ is the maximum external pump intensity. The source term in (\ref{4.11}) is then given by
\begin{equation}
J (\xi) = I_0 \epsilon_\chi e^{- i \kappa_\chi \xi / 2} \exp \left( - \frac{\xi^2}{\sigma_0^2} \right).
\label{4.14} 
\end{equation}
\begin{figure*}[ht!]
\includegraphics[scale=0.7]{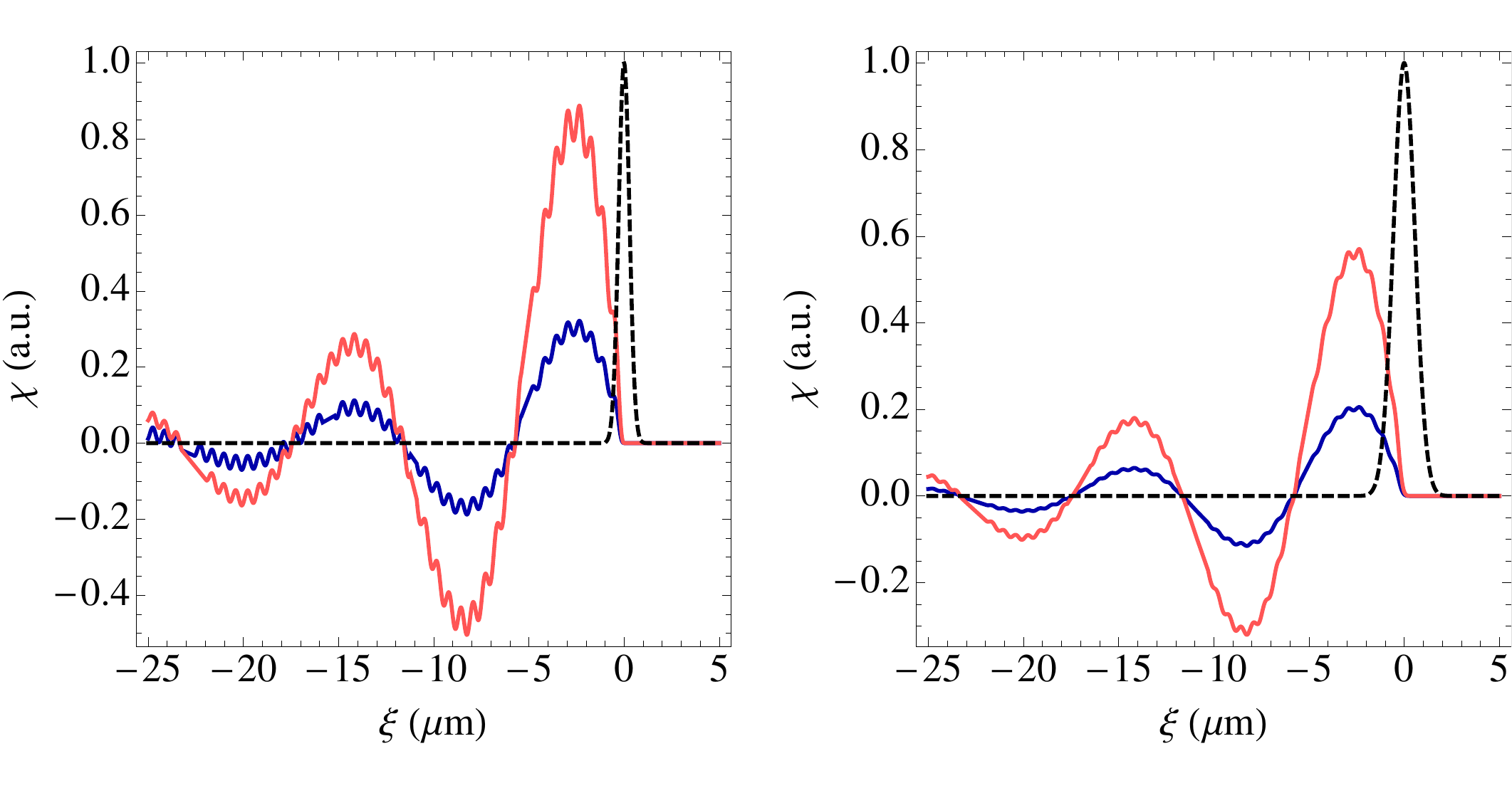}
\caption{(color online) Numerical example of a double exciton-polariton wakefield, produced by a moving external source, corresponding to the following parameters: $m_\phi=5\times 10^{-5}m_e$, $m_\chi=0.4m_e$, $\tau_\phi=30$ ps, $\tau_\chi=400$ ps and $\hbar\Omega_R=10$ meV.  The red/light gray lines are obtained for $v=0.1$ nm/ps, while the blue/darker lines correspond to $v=0.08$ nm/ps. In the left (right) panel, we used $\sigma_0=0.1~\mu$m ($\sigma_0=0.3~\mu$m). We have used the same cavity parameters as in Fig. \ref{fig2}.}
\label{fig4}
\end{figure*}
The polariton wakefield solution is illustrated in Fig \ref{fig4}. We can clearly identify two different scales, one determined by the exciton dynamics, which is similar to the one shown in Fig.\ref{fig2} (slow modulation), and a second one determined by the photon dynamics (fast modulation). We also observe that the fast modulation, which possess smaller amplitude, is less damped that the slow mode. This is so because the exciton lifetime is much larger than the photon mode. We notice the that slow and fast oscillations propagate with the same phase velocity, which equals the velocity $v$ of the moving source. Such a synchronism between the velocity of the moving object and the phase velocity of the wakefield oscillations is a characteristic features of the wakefields \cite{tajima, ali}. One interesting application of wakefields in practical experiments can be the characterization of semi-conductor micro-cavities, as the form of the wake is strictly related on the basic properties of the semi-conductor and the cavity (quality factor, excitonic life-time, coupling strength, etc). The access to time-resolved spectra, with the aid of a streak camera, may allow direct and very accurate measurements of wakefield features (modulation frequencies, amplitude and damping rate).

\section{Conclusions}

We have investigated the generation of wakefields in a semi-conductor microcavity in the strong coupling regime excited by a moving external pump source. This opens the way to the study of non-stationary regimes in driven-dissipative cavities, which could lead to further theoretical studies, and could stimulate a new type experimental observations. We have shown that the nature of the wake, being either purely excitonic or polaritonic, can be controlled by tuning the pump wavevector to become resonant at different positions of the lower polariton dispersion.  While in the former case the wakefield corresponds to a single frequency modulation of the exciton wavefunction envelope, in the latter the wakefield consists of a two-frequency modulation of the polariton envelope. The slow (fast) mode is associated with the excitonic (photonic) fraction of the polariton fluid. We have also argued that the wakefield move with phase velocity coinciding with the velocity of the source. The influence of the transverse shape of the moving source, the relation of the resulting wakefields and the possible Cherenkov emission will be left to a future work.

\begin{acknowledgements}

One of the authors (HT) would like to acknowledge the support of the EU POLAPHEN, ANR Quandyde and GANEX projects.
The other author (JTM) would like to thank the financial support of CAPES Brasil,  and the hospitality of the Institute of Physics of the University of S\~ao Paulo. 

\end{acknowledgements}

\end{document}